\RequirePackage{fix-cm}

\documentclass[smallextended]{svjour3}

\smartqed

\usepackage{graphicx}
\usepackage{array}

\title{Quality model for evaluating and choosing a stream processing framework architecture}

\begin{document}

\author{Youness Dendane \and Fabio Petrillo \and Hamid Mcheick \and Souhail Ben Ali}
\institute{Université du Québec de Chicoutimi\\
            Department of Mathematics and Computer science\\
            555 boulevard de l'Université\\
            Chicoutimi, Canada\\%
	    \email{dendaneys@gmail.com,fabio@petrillo.com,hamid\_mcheick@uqac.ca,souhail.ben-ali1@uqac.ca}            
}

\date{2019 Jan}

\maketitle

\begin{abstract}
Today, we have to deal with many data (Big data) and we need to make decisions by choosing an architectural framework to analyze these data coming from different area. Due to this, it become problematic when we want to process these data, and even more, when it is continuous data. When you want to process some data, you have to first receive it, store it, and then query it. This is what we call Batch Processing. It works well when you process big amount of data, but it finds its limits when you want to get fast (or real-time) processing results, such as financial trades, sensors, user session activity, etc. The solution to this problem is stream processing. Stream processing approach consists of data arriving record by record and rather than storing it, the processing should be done directly. Therefore, direct results are needed with a latency that may vary in real-time. 

In this paper, we propose an assessment quality model to evaluate and choose stream processing frameworks. We describe briefly different architectural frameworks such as Kafka, Spark Streaming and Flink that address the stream processing. Using our quality model, we present a decision tree to support engineers to choose a framework following the quality aspects. Finally, we evaluate our model doing a case study to Twitter and Netflix streaming.

\end{abstract}

\section{Introduction}
More and more data is produced today, and different techniques have been developed in order to process this data. Due to modern Big Data applications, like sensors, stock-trading or even user web traffic \cite{c6} data has to be processed in real-time. The technique that can handle this problem is called : stream processing \cite{c5}.\\
So we have assisted to the rise of Stream processing frameworks, such as Samza and Flink, which are becoming more and more popular, for offering a model to ingest and process data at near real-time \cite{c7}.\\
However, with several stream processing frameworks and technologies associated available, a problem arise : how to choose the right framework ? Each framework has its own features and is more or less different from another framework. \\
So, depending on the context, you choose the best solution. But another problem occurs here : on what criteria are you basing on to answer this question ? \\
In this paper, we provide a quality model for a decision taking. This model enforced by what we call variables/criteria, can help you through a decision and we see if it is suitable to choose stream processing framework. \\
We identify and explain in details four criteria that are important for the framework decision making. Further, we quickly present the selected frameworks with their pros and cons. The criteria and the frameworks have been chosen following a study of stream processing papers. We analyzed these papers, and picked based on an average, the most redundant.\\
The rest of the paper is organized as follow, we analyze the related work that has been done (ii), and then answer to the previous questions by identifying what are the different criteria you have to base (iii) and by introducing the different chosen stream processing frameworks (iv). We propose a decision model tree supported by the previous parts, that you can base on to choose the right framework technology (v). 

\section{State-of-the-art/ Related Work}
A stream processing system requires four major elements: 
(1) Best understanding of the streaming application’s architecture (2) identification of key requirements of distributed stream processing frameworks (DSPF) that can be used to evaluate such a system, (3) survey existing streaming frameworks, (4) evaluation and a comparative study of the most popular streaming platforms. We divide the related work based on the three elements mentioned above.

\subsection{Architecture of streaming applications}
Streaming application’s architecture is not too much different from web architectures.
Streaming sources are communicating using arbitrary protocols. So that, a gateway layer is set up to connect sources to streaming application and resolve the heterogeneity of source’s protocols.
A message queues are set up as a middleware to provide a temporary buffer and a routing layer to match the accepted event sources and the applications \cite{c11}.

\subsection{Requirements of distributed stream processing frameworks}
There are eight rules \cite{c12} that serve to illustrate the necessary features required for any system that will be used for high-volume low-latency stream processing applications.

\begin{itemize} 
\item Rule 1: Keep the Data Moving by achieving a low latency
\item Rule 2: Query using higt level language like SQL on Streams (StreamSQL)
\item Rule 3: Handle Stream Imperfections (Delayed, Missing and Out-of-Order Data)
\item Rule 4: Generate Predictable Outcomes
\item Rule 5: Integrate Stored and Streaming Data
\item Rule 6: Guarantee Data Safety and Availability
\item Rule 7: Partition and Scale Applications Automatically
\item Rule 8: Process and Respond Instantaneously
\end{itemize}

\subsection{Existing streaming frameworks}
Several streaming frameworks have been proposed to allow real-time large scale stream processing. In this section sheds the light on the most popular big data stream processing frameworks:

\subsubsection{Apache Spark \cite{c15}}
Developed at UC Berkeley in 2009 \cite{c19}, is a platform for distributed data processing, written in Java and Scala. In spark, streaming computation is treated as a series of deterministic batch computations on small time intervals.

\subsubsection{Apache Storm \cite{c18}}
is a real-time stream processor, written in Java and Clojure. Storm is a fault tolerant framework that is suitable for real time data analysis, machine learning, sequential and iterative computation.

\subsubsection{Apache Flink \cite{c17}}
is an open source processing framework supporting both stream and batch, It provides several benefits such as fault-tolerant and large scale computation \cite{c14}. Multy functionalities are offred by this plateform such us additional high level functions such as join, filter and aggregation it allows iterative processing and real time computation on stream data collected by different tools such as Flume \cite{c20} and Kafka \cite{c21}.

\subsubsection{Apache Samza \cite{c16}}
is created by Linkedin to solve various kinds of stream processing requirements such as tracking data, service logging of data, and data ingestion pipelines for real time services \cite{c14}. It uses Apache Kafka as a distributed broker for messaging, and Hadoop YARN for distributed resource allocation and scheduling \cite{c14}.

\subsection{A comparative between processing frameworks}
The comparison between those several frameworks listed above are data format, types of data sources, programming model,
cluster manager, supported programming languages, latency and messaging capacities \cite{c14}.

\begin{figure}
  \includegraphics[scale=0.6]{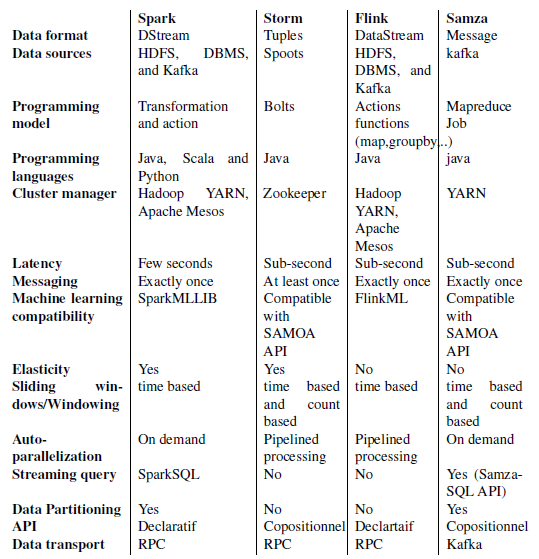}
  \caption{Frameworks comparative}
  \label{fig:comparative}
\end{figure}

\section{Paper Contribution}

The work reported reported in this paper can be categorized under the class of decision help of choosing a stream processing framework. While there is a rich body of work in designing stream processing applications and huge comparative between these applications, a system that can help you to choose the best application by criteria is still messing from contemporary stream processing systems.

In this paper we discuss some architectural frameworks such  as  Storm,  Spark  and  others  that  resolve the  Stream processing  problem and we provide a a  quality  model  to choose  ans  evaluate  a  stream  processing  framework basing  on some criteria such us latency, guarantees, fault tolerance and data processing model.


\section{Survey of Stream Processing Frameworks}
In this section, we will present 4 frameworks that are used actually to resolve stream processing problem.

\subsection{Storm}
Storm integrates with any database (e.g: MongoDB) and any queuing system (e.g: RabbitMQ, Kafka).\\
Storm works with tuples. A tuple is a named list of values and can contain any type of object. \\
Its API is simple and easy to use due to only three abstractions : 

\begin{enumerate}
    \item{Spout : } A spout is a source of streams and reads from a queuing broker.
    \item{Bolt : }Where most of computation's logic goes. Computation logic can be functions, filters, streaming joins, streaming aggregations etc. So basically, from an input, and with computation logic you can produce new output streams.
    \item{Topology : } A network of spouts and bolts.
\end{enumerate}
\par Storm is scalable, fault-tolerant and have an\textbf{ at-least once} guarantee message semantic. The cons here are that there is not ordering guarantees and duplicates may occur. \\
Another of its strengths is if a node dies, the worker will be restarted on another node. If a worker dies, Storm will restart it automatically.\\
At the date of writing this article, with Storm SQL integration, queries can be run over streaming data, but it is still experimental.\\
Furthermore, Storm provides an \textbf{exactly-once} guarantee with Trident which is a high-level abstraction. This model is a micro-batch processing model that add a state and will increase latency.  

\subsection{Spark}
Spark is an hybrid framework which means it can perform batch as well as stream processing.\\
Spark natively works with batch, but it has a library called Spark Streaming that can allow to work with near real time data. It means that incoming data are regrouped into small batch and then processed without increasing the latency too much unlike Storm which provides true streaming processing.\\
One of its power is that the manner you write batch jobs is the same you write stream jobs. More than that, it is fault-tolerant and has an \textbf{exactly-once} semantics.\\
Spark has its own modules that you can combine : 
\begin{itemize}
    \item Spark SQL
    \item Spark Streaming
    \item Machine Learning
    \item GraphX (for graph programming)
\end{itemize}
Spark runs in Hadoop, Apache Mesos, Kubernetes, standalone or in the cloud and access diverse data sources such as HDFS, Cassandra, etc.



\subsection{Samza}
Samza is decoupled in three layers \cite{c8} :  
\begin{enumerate}
    \item Streaming
    \item Execution
    \item Processing
\end{enumerate}

\subsubsection{Streaming}
For the message queuing system, Samza uses Kafka. Kafka is a distributed pub/sub and it has an at-least once message guarantees. Kafka consumers subscribe to topic, which allow them to read messages.

\subsubsection{Execution}
Samza uses YARN to run jobs. It allow to execute commands on a cluster of machines after allocating containers. This is made possible because of YARN, which is the Hadoop's next generation cluster scheduler. So, YARN provides a resource management and task execution framework to execute jobs.

\subsubsection{Processing}
It uses the two layers above; input and output come from Kafka brokers. YARN is used to run a Samza job and supervise the containers. The processing code the developer write runs in these containers. Samza's processing model is real time.\\

One of Samza's advantages is that the streaming and execution layers can be replaced with any other technologies. Also, because of the use of YARN, Samza is fault tolerant; Samza works with YARN to transparently migrate tasks to another machine.\\
The processing model Samza provides are both batch and stream (real time). Whatever the code you write, it will be reusable whatever the model. Switching models needs config change; from HDFS to Kafka to pass from batch to stream processing.  

\subsection{Flink}
Flink supports batch and real-time stream processing model. It has an exactly-once guarantee for both models. Flink is fault-tolerant and can be deployed to numerous resource providers such as YARN, Apache Mesos and Kubernetes; but also as stand-alone cluster.\\
One of the advantages of this framework is that it can run millions of events per seconds by using the minimum of resources, all of this at a low latency. \\
Flink provides three layered API's :
\begin{enumerate}
    \item ProcessFunction :  It implements the logic, process individuals or grouped events and give control over time and state.
    \item DataStream : Provides primitives for stream operations such as transformations. It is based on functions like aggregate, map and reduce.
    \item SQL : To ease the writing jobs for analytics on real time data.
\end{enumerate}

\section{Criteria used in frameworks}
To choose a stream processing framework, we have identified some criteria. These criteria don't give you the answer on whether you should use stream processing or batch processing, but rather helps you take the decision to pick the right framework. So this step assumes that you already identified the problem and you came to the idea that should use stream processing model over batch processing.

We first are going to give the criteria and explain them in details : 
\begin{itemize}
    \item Latency
    \item Message semantics (guarantees) 
    \item Fault tolerance 
    \item Data processing model (micro-batch or real-time)
\end{itemize}

\subsection{Message semantics}
Another term referring to this criteria is \textbf{Message guarantees}. The message guarantees can take three forms : 
\begin{itemize}
    \item At least-once : could be duplicates of the same message but we are sure that it has been delivered
    \item At most-once : the message is delivered zero or one time
    \item Exactly-once : the message is guaranteed to be delivered exactly one and only one time
\end{itemize}

Before providing message guarantees, system should be able to recover from faults. \cite{c6}

\subsection{Fault tolerance}
Streaming application run for an indefinite period, so it increases the chance of having faults. So this criteria is important, because despite the application has faults. \\
Fault tolerance guarantees that the system will be highly available, operates even after failures and has possibility to recover from them transparently. Flink has the highest availability.

\subsection{Latency}
Latency is the time between arrival of new data and its processing \cite{c10}. Latency goes hand in hand with recovery (fault tolerance) because, whenever the system has errors, it should recover fast enough so the latency doesn't decrease too much (i.e : the processing continue with minimal effect). Also, each framework can do do some optimization on data such as message batching, to improve the throughput, but the cost is sacrificing latency.

\subsection{Data processing model}
To do stream processing, there is two techniques :
\begin{itemize}
    \item Micro-batch : based on batch processing but rather than processing data that have been collected over previous time, data is packaged into small batches and collected in a very small time intervals and then delivered directly to the batch processing. Spark for example does micro-batch.
    \item Real-time : data is processed on fly as individual pieces, so there is no waiting. Flink process data in real-time.
\end{itemize}

As messages are received directly the real-time processing technique has a lower stream processing latency than micro-batch but it become harder to have an exactly-once semantics. However, micro-batch provides better fault-tolerance and thus it can guarantees that the message has been received only once (i.e : Spark Streaming). \\

What we understand here is that message semantics are related to the fault tolerance and the data processing model, and according to how the fault tolerance is implemented the latency will increase or decrease. 

\section{Quality Model for choosing and evaluating a SPF}
After presenting the different frameworks and found the main characteristics/criteria, we came with a model. A model for evaluating the frameworks and choosing one given a set of criteria. In this section, we explain why we have chosen these particular frameworks and how we extracted certain criteria. Afterward, we explain how we have prioritized the criteria, and then, with all these information we present the quality model.

\subsection{Methodology}
There is several processing frameworks used in production today. But to find out what framework is used in which company is difficult and take time. So, our primary support was the research papers. We analyzed various papers about stream processing, and we defined \textbf{redundancy} as our benchmark. This means that we made a table with the papers and frameworks, and every time a paper cited a framework we gave a point to the paper. At the end, we had a table with the frameworks cited per paper.

\begin{figure*}
  \includegraphics[width=\textwidth]{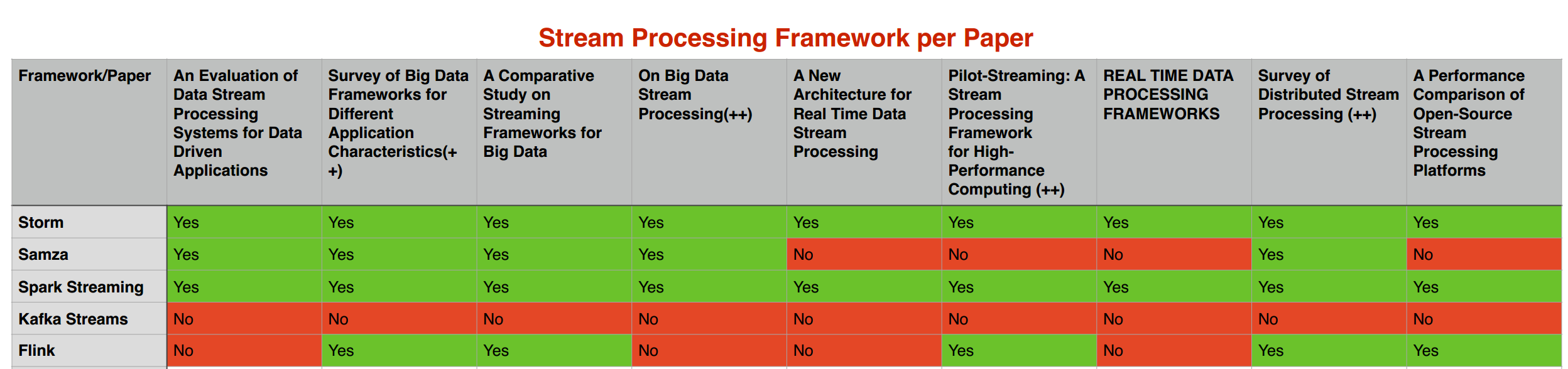}
  \caption{Frameworks per paper}
  \label{fig:frameworks-paper}
\end{figure*}

\begin{figure*}
  \includegraphics[width=\textwidth]{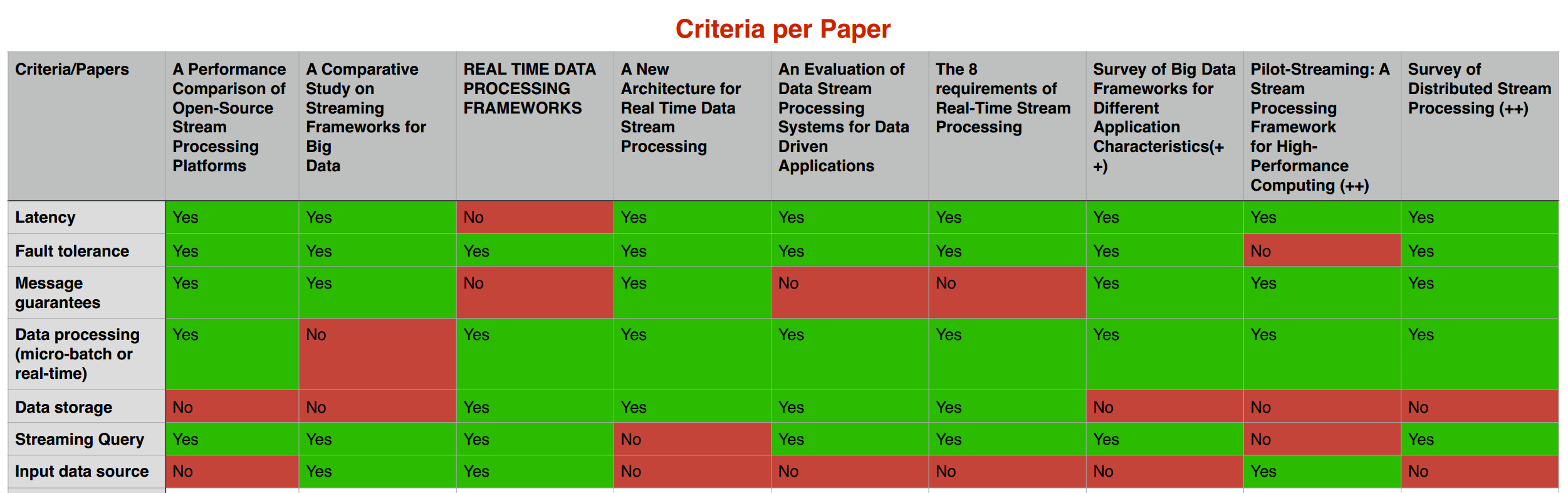}
  \caption{Criteria per paper}
  \label{fig:criteria-paper}
\end{figure*}

We repeated the same process for the criteria. The result is on figure \ref{fig:criteria-paper}.

This paper is a first draft, and we plan to study more papers to have more criteria and frameworks, and thus, to have better average results.

\subsection{Choosing and prioritizing the criteria}
After finding the criteria, we had to prioritize them. Here is the criteria ranked by importance.

\begin{enumerate}
    \item Data model
    \item Fault tolerance
    \item Message semantics
    \item Latency
\end{enumerate}

The first decision is what type of stream processing to choose, because this will have an impact on the other criteria. If you choose a micro-batch framework, it will be possible to have for each framework an exactly-once message semantics as opposite to a real-time model.

Latency is of great importance, but, a framework should be able to recover fast enough, so it does not affect the system too much (with minimum time). And before providing message semantics it also should be recover from faults automatically. Because it will influence the other criteria beneath it, this is why the fault tolerance is in second position.

Depending on whether it is exactly-once or at least-once message semantics, the latency will change depending this criteria.

\subsection{Decision Model Tree}
\begin{figure*}
  \includegraphics[width=\textwidth]{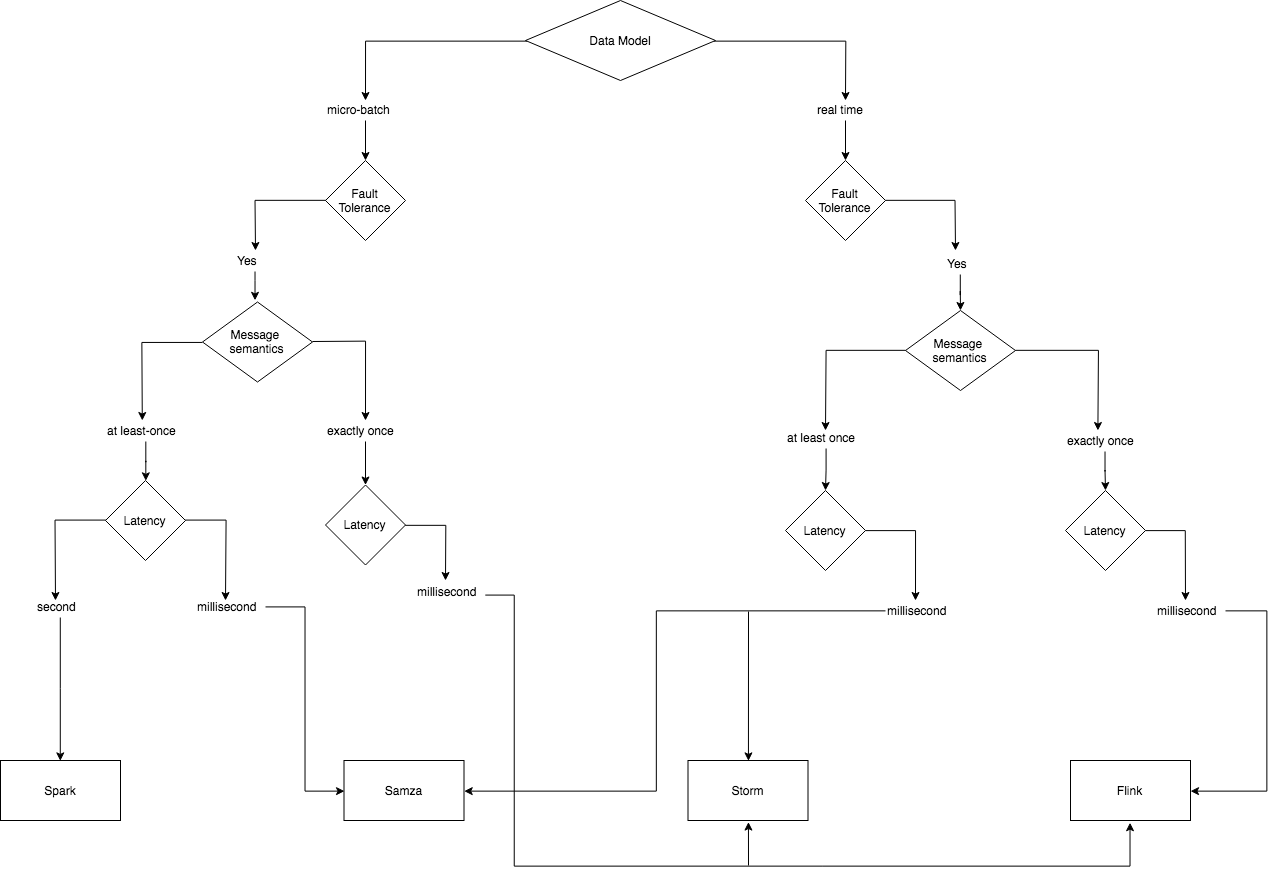}
  \caption{The decision model tree}
  \label{fig:decision-model}
\end{figure*}

Based on the previous parts, we present the decision model tree to evaluate and choose a stream processing framework (fig. \ref{fig:decision-model}). 

\section{Case studies}
In this section, we analyze some stream processing application cases. We go through two companies : Netflix and Twitter. \\
The goal of this section is to see if our contribution in this paper correspond to the reality (i.e: real world application). In analyzing how and why these companies use stream processing frameworks, we can identify the main underlying elements and compare them to our criteria. We get all information from papers and the companies tech blog.

\subsection{Twitter}
Twitter has actually an in-house framework called Heron. But before that, they were using Storm. We are going to detail framework evaluation for Storm, because Heron is an improvement but they are still using what we detail below.

The company that has made Storm was acquired by Twitter in 2011. Since, Twitter modified for their use.\\ 
Let's begin with our first criteria : data processing model.
At Twitter, due to choosing Storm, as we described it above, it has a micro-batch processing model. So, just by using it, the choice of data processing model has been made. 
We go now to our second criteria : fault tolerance. When Twitter describes Storm \cite{c18}, they say that one of the argument chosen to design Storm is : resilient (i.e : fault tolerant); their second criteria and ours correspond. As they say in the article \cite{c18}, on of the feature key is the processing semantics or message semantics. They describe that their solution has two guarantees : at least once and at most once. This characteristic correspond to our third criteria we have mentioned. Further in the article, Ankit et al. report some experiment they have made that had to show the latency results. As they calculated, their latency is close to 1ms 99\% of the time. Our criteria are justified by the design and the use of Storm at Twitter. 

In this first subsection, we can conclude that our criteria are match with the main characteristics of design and use of Storm at Twitter.

\subsection{Netflix}
In their article \cite{c22}, they describe Keystone which is their stream processing platform. The solution chosen to do stream processing is Apache Flink. By choosing Flink, they automatically chosen the real-time processing for the data model criteria. Then, they gave a summary of common asks and trade-offs and one of them is failure recovery. This correspond with our criteria. One of the asks was that the system is fault tolerant. If we follow our model, the next step is to choose the message semantics. In the post, their say that according to the use case loosing some events in the pipeline is acceptable while in other cases the event have to absolutely processed so it require a better durability. We see that this sentence is a synonym to our message guarantees criteria. In another post \cite{c23}, they describe this time a real use case : to know what is trending on Netflix. In order to that, they need real-time data of what users watch, the event is then send to be processed. They describe that one of their challenges was having a low latency. This last criteria match with ours.

What we can conclude in this section is that these companies followed a path which correspond with our quality model. All our criteria had been taken into account by these companies and are part of the core decision on choosing and using stream processing framework architecture.

\section{Discussion}
In this section we will discuss the impact of our results, impact as well on engineers as on researchers. This quality model can be used as a guideline when wanting to choose a stream processing framework. Answering what type of criteria is important for a given context will end to the choice of the right solution; do I need absolutely only one instance of data or is it permissible to have duplicates ?  (i.e: at least once vs exactly once semantics). Answering to these questions based on the criteria we identified will help the engineers make the right choice quicker. Further, the use case of our model is not limited to the choice only. Our model can be extended to serve to design a future stream processing framework architecture. When designing the solution, the model can help to see further steps on what will be implemented and thus the different dependencies it will have : when implementing the fault tolerance, the latency will increase or decrease given on how it is implemented. More over, thanks to the model, we see that the fault tolerance will also influence the message semantics. So based on what we want to have as message guarantees, we will implement the fault tolerance in a different manner. In the other hand, researchers can use this model when wanting to evaluate a framework architecture. Also, this model, can be reused in order to compare different frameworks. When wanted, as part of their research, they can have a quicker and a better view on the different solution and what brings to them and how they are different and also similar. More over, when wanted and depending on their need, they can easily extend this quality model in order to adapt it to their work : adding a criteria will add complexity, and thus a possible different path.

\section{Conclusion \& Future work}
With the huge amount of data generated, and given a stream processing context, choosing the right framework architecture is major. In order to do that, we first identified and explained what are the different criteria such as data model and latency... and presented some stream processing frameworks. We explained our methodology on how we came to choose the ideal framework architecture to fulfill user's needs. Given these, we provided a decision model tree which is a quality model to choose and evaluate a stream processing framework.\\
There is more work that has to be done, in order to have more criteria and frameworks, thus to have a more complete and complex model. We can base on this model to evaluate and choose a framework architecture, and not only that, this model can also serve as a guide to designing a new stream processing framework architecture. It can also be used as a support to have quickly a global view of the different solution and what brings to them depending on the different criteria.

\addtolength{\textheight}{-12cm}   




\end{document}